\documentclass[multphys,vecphys]{svmult}

\usepackage{makeidx}         
\usepackage{graphicx}        
\usepackage{multicol}        
\usepackage[bottom]{footmisc}

\makeindex             

\begin{document}

\title*{Planets around active stars}
\titlerunning{Planets around active stars}
\author{J. Setiawan\inst{1}\and P. Weise\inst{1}\and Th. Henning\inst{1}\and A.P. Hatzes\inst{2}\and L. Pasquini\inst{3}\and 
L. da Silva\inst{4}\and L. Girardi\inst{5}\and O. von der L\"uhe\inst{6}\and M.P. D\"ollinger\inst{3}\and A. Weiss\inst{7}\and 
K. Biazzo\inst{3,8}} 
\authorrunning{J. Setiawan et al.}
\institute{
Max-Planck-Institut f\"ur Astronomie, Heidelberg, Germany setiawan@mpia.de
\and Th\"uringer Landessternwarte, Tautenburg, Germany
\and European Southern Observatory, Garching bei M\"unchen, Germany
\and Observat\'orio Nacional, Rio da Janeiro, Brazil
\and Osservatorio Astronomico di Padova-INAF, Padova, Italy 
\and Kiepenheuer-Institut f\"ur Sonnenphysik, Freiburg (Brsg), Germany
\and Max-Planck-Institut f\"ur Astrophysik, Garching bei M\"unchen, Germany 
\and Osservatorio Astrofisico di Catania-INAF, Catania, Italy
}
%
%
\maketitle

\begin{abstract}
We present the results of radial velocity measurements of two samples of active stars. The first sample contains field 
G and K giants across the Red Giant Branch, whereas the second sample consists of nearby young stars (d $<$ 150 pc) with ages 
between 10 - 300 Myrs. The radial velocity monitoring program has been carried out with FEROS at 1.52 m ESO telescope (1999 - 2002) 
and continued since 2003 at 2.2 m MPG/ESO telescope. We observed stellar radial velocity variations which originate either from 
the stellar activity or the presence of stellar/substellar companions. By means of a bisector technique we are able to distinguish 
the sources of the radial velocity variation. Among them we found few candidates of planetary companions, both of young stars and 
G-K giants sample.
\end{abstract}

\section{Introduction}

Precise radial velocity (RV) technique by using high-resolution spectrographs has been very successful in detecting extrasolar planets 
around inactive solar-like stars. Recently, this method has also some success in planet detections around several active stars, e.g., in 
G and K giants [2],[4],[5],[8],[10]. However, the number of planets discovered around such active stars is still very low, compared to 
those of solar-like stars. This is due to the fact, that either the stellar activity, like fast stellar rotation, prohibits accurate RV 
determination or the measured RV variation is indeed not caused by the presence of low-mass companions, instead of by stellar pulsations 
or starspots. 

Therefore, detailed investigations of spectral line profiles by the bisector and other activity indicators are mandatory 
in order to avoid false detections. Beside G-K giants, young stars also belong to stars which are still not well explored with the precise 
RV technique because of their high activity level. However, for less active (slow rotating F, G, K) young stars RV method is still applicable 
and will be able to find close orbiting giant planets which cannot be detected by the current direct imaging methods. The detection 
of young planetary systems will certainly improve our understanding of the planet formation.

In this paper we present results of our RV survey, which include the stability performance of FEROS, detection of stellar activity 
(non-radial pulsations and rotational modulation) and detections of a planet around a young star HD 70573 (t$\sim$100 Myrs) and a 
second planet candidate of an old metal poor K giant star HD 47536 ($t\sim$9 Gyrs, [Fe/H]=$-$0.68).

\section{Observations}

Since 1999 we have been carrying out RV monitoring program of G-K giants with FEROS at 1.52 m ESO in La Silla observatory. The results 
were presented in [7] and [8]. After FEROS was moved to 2.2 m MPG/ESO telescope follow-up observations of several G-K giant targets 
with long-period RV variation have been done. In 2003 we added nearby young stars ($t =10-300$ Myrs) to our FEROS sample to be observed 
within the MPG time. To our surprise, about 30\% of young stars in the sample show low-level stellar activity and slow stellar rotation, 
making them suitable for accurate RV measurements. The accuracy of FEROS at 2.2 m MPG/ESO from Dec 2003 until Oct 2006 is 10 m/s (Fig. 1), 
which is a significant improvement than at 1.52 m ESO.

\begin{figure}
\centering
\includegraphics[width=5.cm,height=4.3cm]{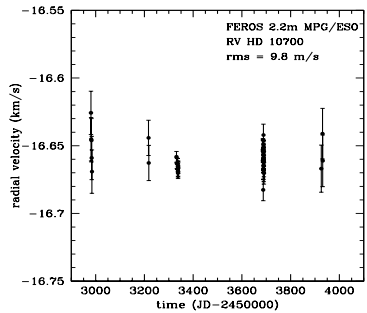}
\caption{RV measurements of a standard star (HD 10700) from Dec 2003 - Oct 2006 with FEROS at 2.2 m MPG/ESO telescope, La Silla.}
\end{figure}

We used a cross correlation technique to measure the stellar RVs. Basic stellar parameters ($M$, $T_{\rm eff}$, [Fe/H], $\log g$ and $R$) 
for our G-K giant targets have been determined from the spectroscopic analysis [1]. For the young stars sample we are currently working on 
the method for the stellar parameters measurement.

\section{Stellar activity}

From our RV survey we found short-period and long-period RV variations in both samples. To distinguish intrinsic stellar activity, e.g. 
non-radial pulsations or stellar rotational modulation due to starspots, from the signature of substellar companion we used the bisector 
technique [3],[12]. In Fig. 2 we show an example of a star from our young star target list, where the bisector velocity spans correlate 
with the RVs. Thus, the RV variation is more likely due to the stellar activity rather than the presence of a substellar companion.

\begin{figure}
\centering
\includegraphics[width=4.2cm]{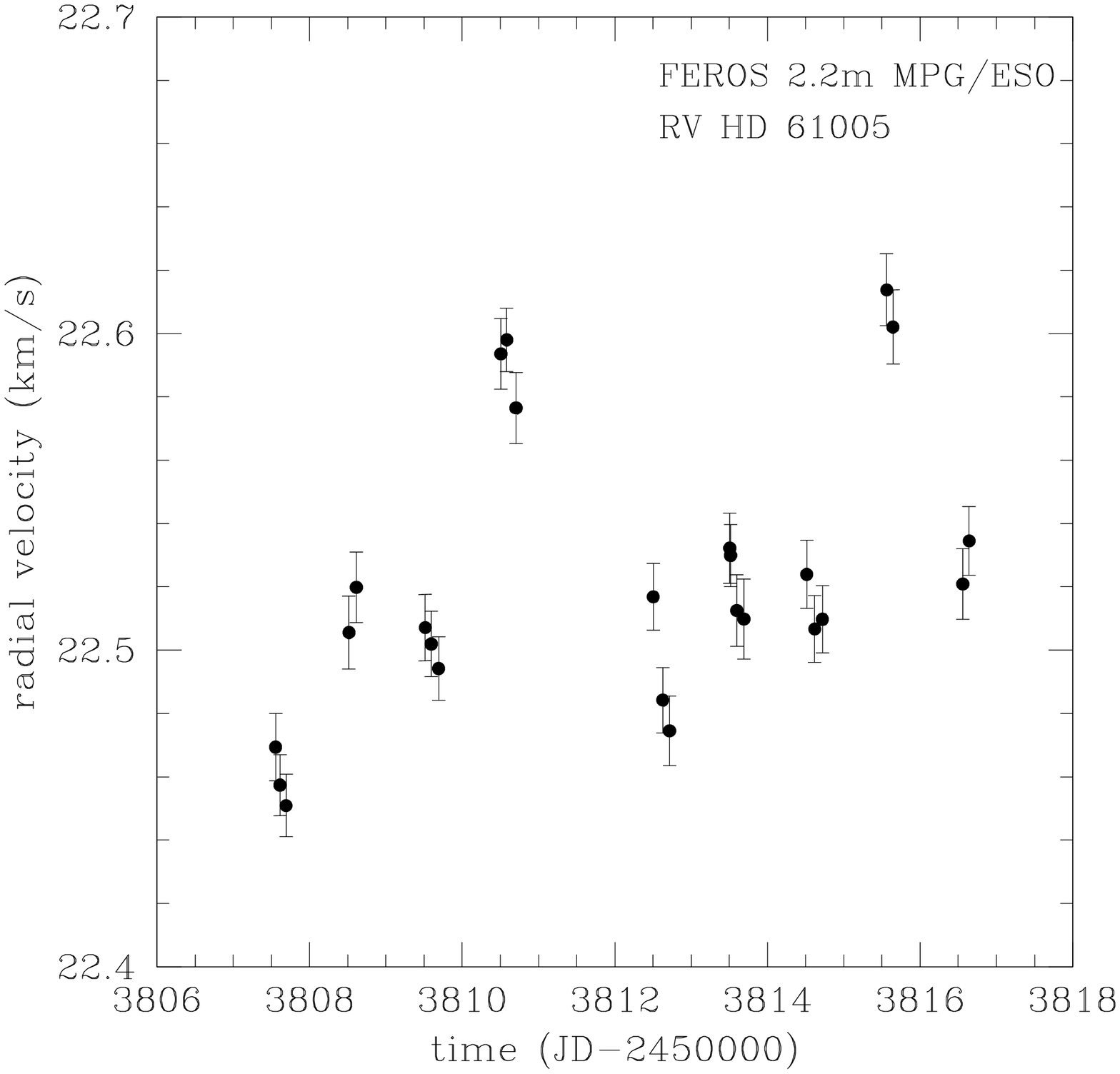}
\includegraphics[width=4.2cm]{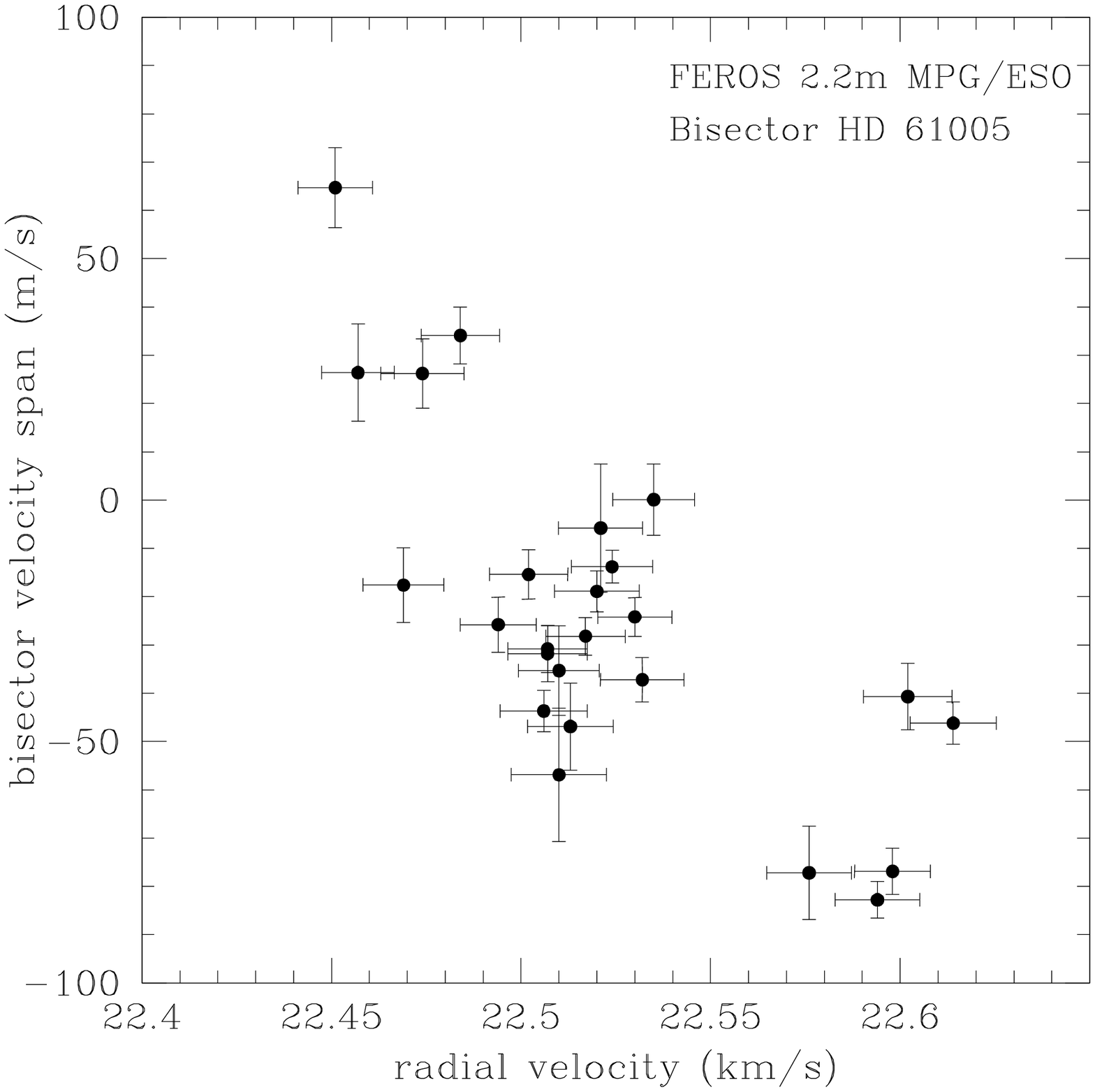}
\caption{RV measurements of a young star HD 61005 (left). The right panel shows a correlation between the bisector and RV. Thus, the 
source of the RV variation is possibly the stellar activity.}
\end{figure}

In the same way, non-planet detections (starspots and pulsations) in K giants have been found in HD 78647 [9] and HD 81797 [11].

\section{Planetary companions}

Fig. 3 (left panel) shows RV measurements of a nearby young star HD 70573 (d = 46 pc, t$\sim$100 Myrs). The RV variation of this star 
shows a periodicity of 850 days which is significantly longer than the stellar rotational period of 3.296 days [6]. Thus, rotational 
modulation as the source of RV variationis unlikely. Moreover, we found no correlation between bisectors and RVs. We also did not find 
any correlation between the RVs and variation in CaII K emission lines. We concluded, that the RV variation is caused by the presence 
of a substellar companion. Assuming a one solar-mass for HD 70573,we calculated a companion's minimum mass of $\sim$6 $M_{\rm Jup}$ and 
a preliminary orbit with a semi-major axis of $\sim$ 1.8 AU and an eccentricity of e = 0.4 (Setiawan et al. in preparation).

One particular star of our G-K giants sample is HD 47536, which has been detected to harbour a planetary companion [8]. Further spectroscopic 
analysis [1] yields a primary mass of 0.94 $M_{\rm Sun}$ and a sub-solar metallicity [Fe/H]= $-$0.68. Together with HD 13189 [5], HD 47536 belongs 
to the most metal poor stars, which have planetary companions. 

To our surprise, further RV follow-up observations with Coralie and FEROS 
at 2.2 m MPG/ESO (Fig. 3, right panel) show an evidence for another planet candidate in the system. The second planet has an orbit with a 
longer period than the first inner planet. According to this detection, a revision of the orbital solution in [8] has been made. We obtained 
an orbital period of $\sim$430 days for the inner planet and, $\sim$ 2500 days for the outer planet. From the revised orbital solution we calculated 
companion's minimum masses of $\sim$5 $M_{\rm Jup}$ and $\sim$7 $M_{\rm Jup}$. More detailed analysis will be presented in upcoming papers 
(Setiawan et al, in preparation).

\begin{figure}
\centering
\includegraphics[width=4.2cm]{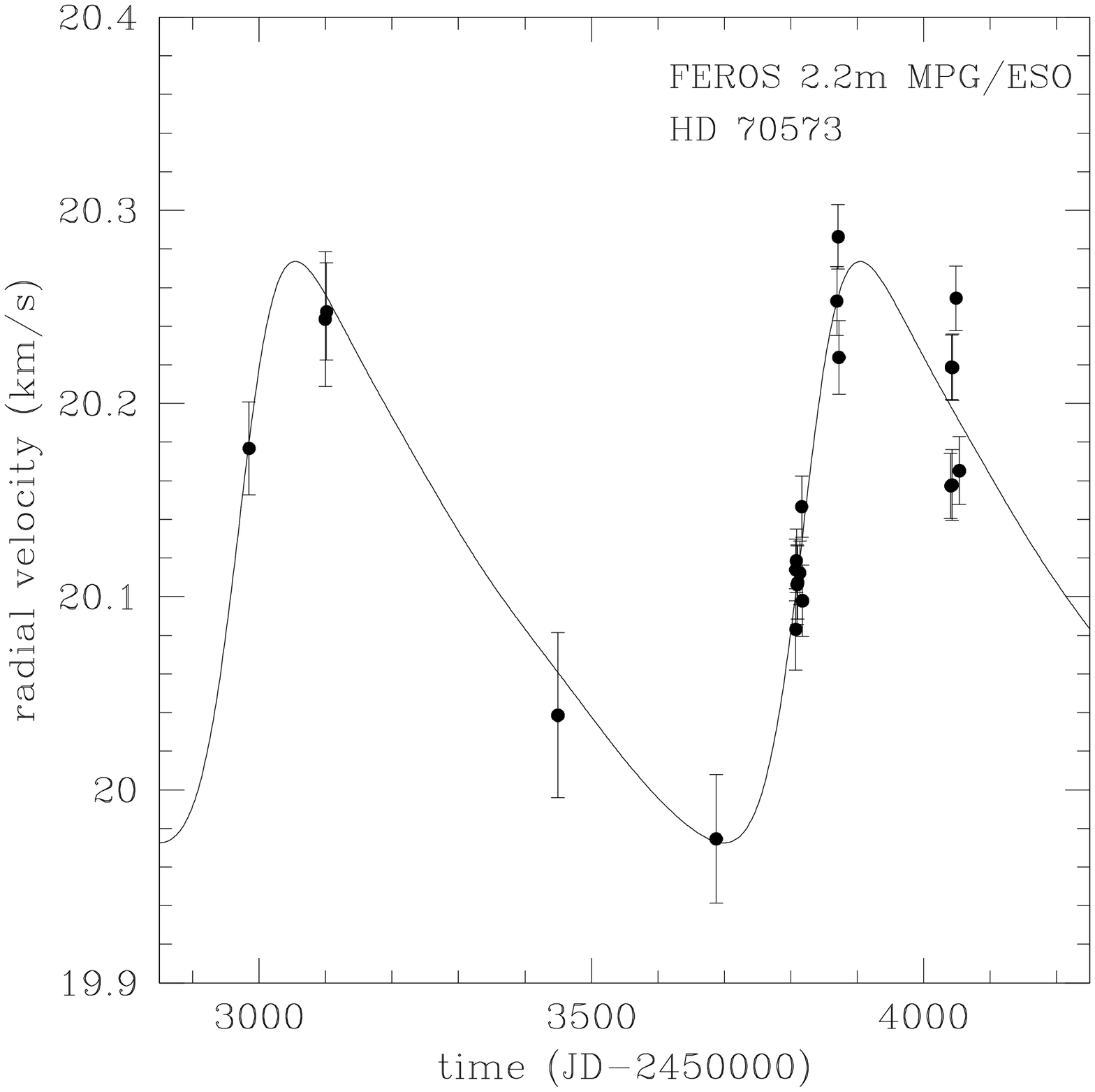}
\includegraphics[width=4.2cm]{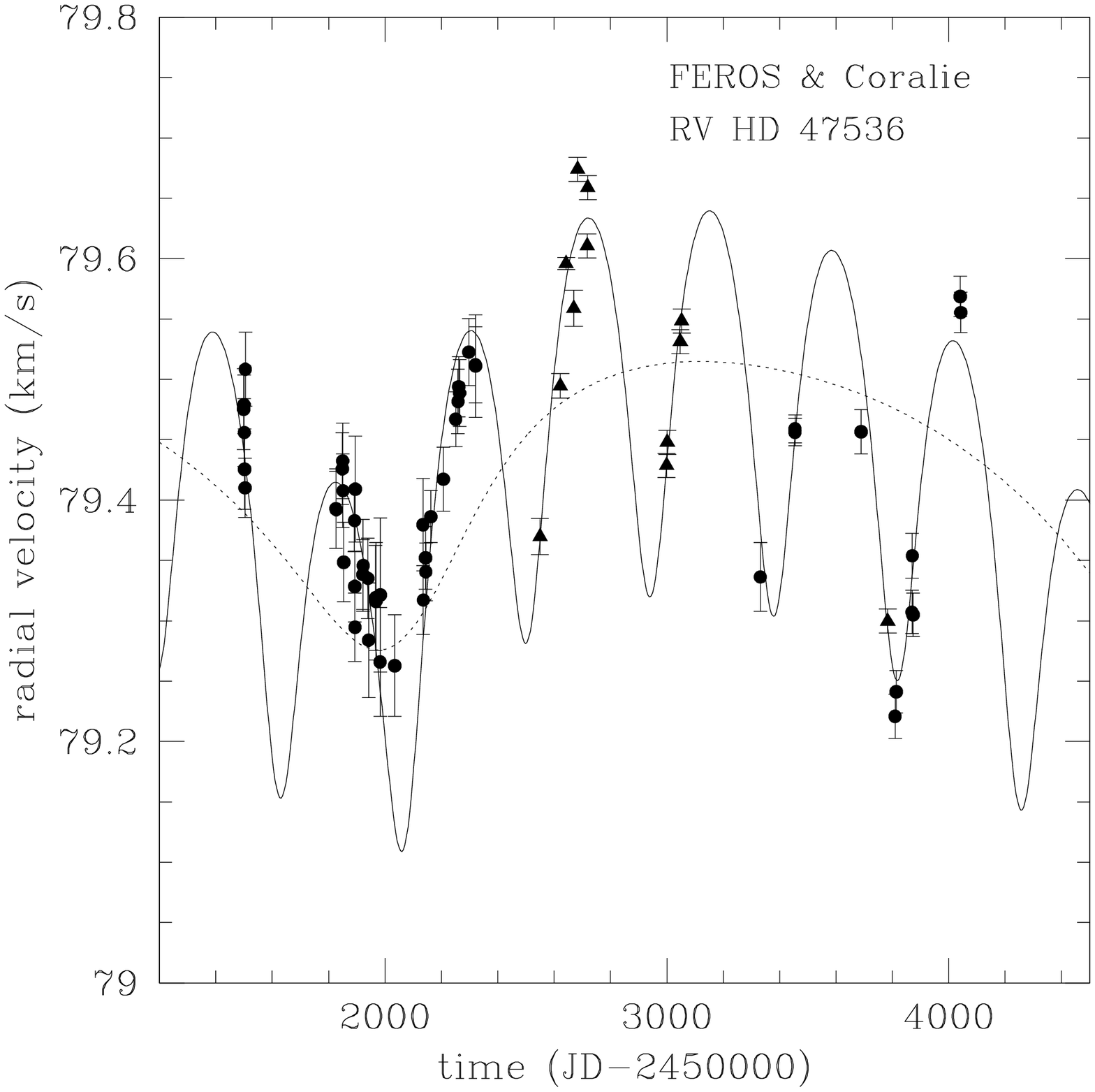}
\caption{Evidence for a planetary companion around a young star HD 70573 (left panel). Right panel: RV measurements of HD 47536 (see text), dots: 
FEROS measurements, triangle: CORALIE measurements.}
\end{figure}

%
%

%
%

\end{document}